# Scaling Laws for Thin Films near the Superconducting-to-Insulating Transition

Yong Tao[†]

College of Economics and Management, Southwest University, Chongqing, China

**Abstract:** We propose a Lagrangian function, which combines Landau-Ginzburg term and Chern-Simons term, for describing the competition between disorder and superconductivity. To describe the normal-to-superconducting transition in the thin superconducting films, we apply Wilson's renormalization group methods into this Lagrangian function. Finally, we obtain a scaling law between critical temperature ($T_c$), film thickness ($d$), sheet resistance of the film at the normal state ($R_s$), and number density of the electrons at the normal state ($N$). Such a scaling law is in agreement with recent experimental investigations [Ivry, Y. et al, Physical Review B 90, 214515 (2014)]. Our finding may have potential benefits for improving transition temperature $T_c$.

Exploring the mechanism of triggering high-$T_c$ superconductivity is an important task. It has attracted many attentions over the past 30 years, and causes a lot of works. Even so, the mechanism of triggering high-$T_c$ superconductivity still remains largely a mystery[1]. However, thin superconducting films are believed to facilitate the comprehension of high-$T_c$ superconductivity, since they allow an inherent competition between disorder and superconductivity which in turn enables the intriguing superconducting-to-insulating transition. Recently, by analyzing the data published over the past 46 years for different materials, Ivry et al discovered a universal relationship[2] between critical temperature ($T_c$), film thickness ($d$), and sheet resistance of the film at the normal state ($R_s$):

$$d \cdot T_c = A \cdot R_s^{-B}. \qquad (1)$$

where, $A$ and $B$ are fitting parameters, and $B \approx 0.9$ to $1.1$.

Although the scaling formula (1) may provide a theoretical basis for improving the transition temperature $T_c$, Ivry et al[2] also stressed that this formula cannot be derived by using existing theories[3–6]. The main purpose of this paper is to investigate the possible origin of the formula (1) by using Wilson's renormalization group methods.

---

[†] Corresponding author.
E-mail address: taoyingyong@yahoo.com

## Competition between disorder and superconductivity

As is well known, renormalization group is a powerful tool of dealing with critical phenomena. The scaling formula (1) associates the transition temperature $T_c$ with the sheet resistance of the film at the normal state $R_s$. To apply the renormalization group methods, it is natural to seek a Lagrangian function, in which $R_s$ and $T_c$ will emerge as the possible coefficients, for describing the competition between disorder and superconductivity. To this end, let us consider superconducting state and normal state, respectively.

(i). *Superconducting state*. As is well known, the superconducting state can be described by the Landau-Ginzburg Lagrangian function in 3-dimensional space:

$$L_{super} = \left|(\partial_\mu - iqA_\mu)\phi\right|^2 + \lambda_2|\phi|^2 + \lambda_4|\phi|^4 + \frac{1}{4}F_{\mu\nu}F_{\mu\nu}, \qquad (2)$$

where, $A_\mu$ denotes the electromagnetic potential, $\phi$ denotes the order parameter, and $F_{\mu\nu} = \partial_\mu A_\nu - \partial_\nu A_\mu$.

(ii). *Normal state*. To describe the normal state we propose to consider the following Lagrangian function in 3-dimensional space:

$$L_{norm} = \frac{\sigma}{2}\varepsilon^{\mu\nu\lambda}A_\mu\partial_\nu A_\lambda + A_\mu J^\mu, \qquad (3)$$

where, $\varepsilon^{\mu\nu\lambda}$ denotes the totally antisymmetric symbol in 3-dimensional space, and $J^\mu$ denotes the current density.

Applying Euler-Lagrange variational procedure into the Lagrangian function (3) yields:

$$J^\mu = -\sigma \cdot \varepsilon^{\mu\nu\lambda}\partial_\nu A_\lambda. \qquad (4)$$

In 2+1 dimensional space-time the coefficient $\sigma$ of Chern-Simons term plays the role with order parameter[7] and has the physical meaning of Hall conductivity[7-9]. In fact, for 2+1 dimensional space-time the equation (4) reproduces Ohm-Hall law[10] in which the coefficient $\sigma$ stands for Hall conductivity. It is well known that Ohm's law describes the normal state well. We hope that the equation (4) for 3-dimensional space can be also used to describe the normal state. Remarkably, Hill[11] has pointed out that the coefficient $\sigma$ in 3-dimensional space must approach the coefficient of Chern-Simons term in 2+1 dimensional space-time; otherwise, there will be inconsistencies (see page 5 in reference 11). Therefore, the coefficient $\sigma$ in 3-dimensional space can still be thought of as the Hall conductivity.

On the other hand, because this paper considers 3-dimensional space rather than 2+1 dimensional space-time, we do not worry that the Chern-Simons term will break time reversal symmetry. Regarding the existence of 3-dimensional Chern-Simons term, we adopt Hill's argument[11]: "*D*-odd Chern-Simons term in a $U(1)$ theory exists at all and what it is physically measuring, i.e., the world-line intersection of Dirac branes."

If we consider the limit $\omega\tau \gg 1$, the resistivity $\rho$ can be written as (see page 1 in reference 12):

$$\rho = \frac{1}{\sigma}, \tag{5}$$

where $\omega$ denotes the angular frequency of the cyclotron motion and $\tau$ denotes the scattering relaxation time.

To describe the competition between disorder and superconductivity, we expect that the Lagrangian functions (2) and (3) can hold simultaneously in the neighborhood of the superconducting-to-insulating transition point $T_c$. Without loss of generality, we assume that the collective behaviors of electrons around the critical point $T_c$ can be described by the following Lagrangian function:

$$L = \left|\left(\partial_\mu - iqA_\mu\right)\phi\right|^2 + \lambda_2|\phi|^2 + \lambda_4|\phi|^4 + \frac{1}{4}F_{\mu\nu}F_{\mu\nu} + \frac{\sigma}{2}\varepsilon^{\mu\nu\lambda}A_\mu\partial_\nu A_\lambda. \tag{6}$$

The Lagrangian function (6) will be the starting point of this paper. It is carefully noted that by our assumption the Lagrangian function (6) only holds in the infinitesimal neighborhood of $T_c$, since the Lagrangian functions (2) and (3) may be invalid in insulating state and superconducting state, respectively. Therefore, the Lagrangian function (6) indeed implies the competition between disorder (non-zero $\rho$) and superconductivity (non-zero $\phi$). However, we need to clarify: we do not investigate what drives the superconducting-to-insulating transition. In fact, we are only interested in what critical behavior will occur around the transition point. From this meaning, our research is merely a phenomenological work.

## Renormalization-group equations

Renormalization group approach is a powerful tool of dealing with critical behaviors. We hope that this approach can reproduce the scaling formula (1). To this end, let us apply renormalization group approach into the Lagrangian function (6).

Adopting the standard procedure of renormalization group we first write down the path integral (see page 396 in reference 13):

$$Z = \int D\phi \int D\phi^* \int DA_\mu e^{-\int d^D x L}, \tag{7}$$

where $D$ denotes the number of space dimensions.

Let us define

$$\phi = \phi_s + \phi_w \tag{8}$$

and

$$A_\mu = A_\mu^s + A_\mu^w, \tag{9}$$

where "$s$" stands for "smooth" part and "$w$" stands for "wriggly" part.

Substituting equations (8) and (9) into equation (7) yields:

$$Z = \int D\phi_s \int D\phi_s^* \int DA_\mu^s \int D\phi_w \int D\phi_w^* \int DA_\mu^w e^{-\int d^D xL}. \tag{10}$$

If we do the integral over $\phi_w$ and $A_\mu^w$, then equation (10) can be rewritten in the form:

$$Z = \int D\phi_s \int D\phi_s^* \int DA_\mu^s e^{-\int d^D xL_{eff}}, \tag{11}$$

where,

$$\int d^D xL_{eff} = \int d^D x \left\{ (1+\Delta E)\cdot \left|(\partial_\mu - i(q+\Delta q)A_\mu^s)\cdot \phi_s\right|^2 + (\lambda_2 + \Delta\lambda_2)\cdot |\phi_s|^2 \right. \\
\left. + (\lambda_4 + \Delta\lambda_4)\cdot |\phi_s|^4 + \frac{1}{4}(1+\Delta F)F_{\mu\nu}^s F_{\mu\nu}^s + \frac{1}{2}(\sigma+\Delta\sigma)\varepsilon^{\mu\nu\lambda}A_\mu^s \partial_\nu A_\lambda^s \right\}. \tag{12}$$

Here $\Delta\lambda_2$, $\Delta\lambda_4$, $\Delta E$, $\Delta q$, $\Delta\sigma$ and $\Delta F$ stand for perturbative terms. Regarding the origin of perturbative terms $\Delta\lambda_2$, $\Delta\lambda_4$ and $\Delta E$, we adopt Peskin's derivation, refer to equation (12.20) in reference 13.

In terms of the rescaled variable $x' = bx$, equation (12) becomes

$$\int d^D x'L_{eff}' = \int d^D x' \left\{ (1+\Delta E)b^{2-D}\cdot \left|(\partial_\mu' - i(q+\Delta q)A_\mu^{s'})\cdot \phi_s\right|^2 + (\lambda_2 + \Delta\lambda_2)b^{-D}\cdot |\phi_s|^2 \right. \\
\left. + (\lambda_4 + \Delta\lambda_4)b^{-D}\cdot |\phi_s|^4 + \frac{1}{4}(1+\Delta F)b^{4-D}F_{\mu\nu}^{s'}F_{\mu\nu}^{s'} + \frac{1}{2}(\sigma+\Delta\sigma)b^{3-D}\varepsilon^{\mu\nu\lambda}A_\mu^{s'}\partial_\nu A_\lambda^{s'} \right\}, \tag{13}$$

where we have used $\partial_\mu' = \frac{\partial x^\nu}{\partial x^{\mu'}}\partial_\nu$ and $A_\mu' = \frac{\partial x^\nu}{\partial x^{\mu'}}A_\nu$.

If we rescale the fields $\phi_s$ and $A_\mu^s$ according to:

$$\phi_s' = \left[b^{2-D}(1+\Delta E)\right]^{\frac{1}{2}}\cdot \phi_s$$

and

$$A_\mu^{s''} = \left[b^{4-D}(1+\Delta F)\right]^{\frac{1}{2}}\cdot A_\mu^{s'},$$

then equation (13) can be rewritten in the form:

$$\int d^D x'L_{eff}' = \int d^D x' \left\{ \left|(\partial_\mu' - iq'A_\mu^{s''})\cdot \phi_s'\right|^2 + \lambda_2'\cdot |\phi_s'|^2 \right. \\
\left. + \lambda_4'\cdot |\phi_s'|^4 + \frac{1}{4}F_{\mu\nu}^{s''}F_{\mu\nu}^{s''} + \frac{1}{2}\sigma'\varepsilon^{\mu\nu\lambda}A_\mu^{s''}\partial_\nu A_\lambda^{s''} \right\}, \tag{14}$$

which shares the same form with Lagrangian function (6).

The new parameters of the Lagrangian function (14) are:

$$q' = (q+\Delta q)(1+\Delta F)^{-\frac{1}{2}}b^{\frac{D-4}{2}}, \tag{15}$$

$$\lambda_2' = (\lambda_2 + \Delta\lambda_2)(1+\Delta E)^{-1}b^{-2}, \tag{16}$$

$$\lambda_4' = (\lambda_4 + \Delta\lambda_4)(1+\Delta E)^{-2}b^{D-4}, \tag{17}$$

$$\sigma' = (\sigma + \Delta\sigma)(1+\Delta F)^{-1}b^{-1}. \tag{18}$$

The attention of this paper will be concentrated on the equations (17) and (18). All of the corrections, $\Delta\lambda_4$, $\Delta E$, $\Delta\sigma$, $\Delta F$ and so on, should be small compared to the leading terms if perturbation theory is justified. This paper only considers the simplest case where the corrections will be ignored. Therefore, equations (17) and (18) can be rewritten as:

$$\lambda_4' = \lambda_4 \cdot b^{D-4}, \tag{19}$$

$$\sigma' = \sigma \cdot b^{-1}. \tag{20}$$

Substituting $b = 1 + \frac{\delta l}{l}$ into equations (19) and (20) we easily get the first-order renormalization group equations:

$$l\frac{d\lambda_4}{dl} = (D-4)\lambda_4, \tag{21}$$

$$l\frac{d\sigma}{dl} = -\sigma. \tag{22}$$

where $l$ is a length somewhat larger than atomic dimensions[14].

## Results

Let us solve the scaling equations from the renormalization-group equations (21) and (22):

$$\lambda_4 = \lambda_0 l^{D-4}, \tag{23}$$

$$\sigma = \sigma_0 l^{-1}, \tag{24}$$

where $\lambda_0$ and $\sigma_0$ are undetermined parameters.
Substituting equations (5) and (23) into equation (24) we get:

$$\lambda_4^{\frac{1}{4-D}} \cdot \rho = \frac{\lambda_0^{\frac{1}{4-D}}}{\sigma_0}, \tag{25}$$

By Bardeen-Cooper-Schrieffer (BCS) Hamiltonian of superconductivity Gorkov[15] has shown that the Landau-Ginzburg equations in 3-dimensional space can be written in the form:

$$\frac{1}{2m^*}(\nabla_{\mathbf{r}} - i2e\mathbf{A}(\mathbf{r}))^2\psi(\mathbf{r}) - \frac{1}{\lambda}\cdot\frac{T-T_c}{T_c}\psi(\mathbf{r}) - \frac{2}{\lambda N}|\psi(\mathbf{r})|^2\psi(\mathbf{r}) = 0. \tag{26}$$

which by rescaling $\psi(\mathbf{r})$ according to $\phi(\mathbf{r}) = \frac{1}{\sqrt{2m^*}}\psi(\mathbf{r})$ yields the following Lagrangian function:

$$L = |(\partial_\mu - i2eA_\mu)\phi|^2 + \frac{2m^*}{\lambda}\cdot\frac{T-T_c}{T_c}|\phi|^2 + \frac{4m^{*2}}{\lambda N}|\phi|^4, \tag{27}$$

where $\lambda = \frac{7\zeta(3)\varepsilon_F}{12\pi^2 T_c^2}$ ($\zeta(x)$ is Riemann's zeta function and $\varepsilon_F$ denotes Fermi energy) and $N$ is number density of the electrons at the normal state[15].

Comparing Lagrangian functions (6) and (27) we obtain:

$$\lambda_2 = \frac{2m^*}{\lambda} \cdot \frac{T - T_c}{T_c}, \tag{28}$$

$$\lambda_4 = \frac{4m^{*2}}{\lambda N}. \tag{29}$$

Comparison of Lagrangian functions (6) and (27) is a key point of obtaining the main result of this paper. Then, a question may arise: "Lagrangian functions (6) and (27) cannot be considered equivalent because Lagrangian function (27) is derived by BCS Hamiltonian, which does not include the Chern-Simons term." However, we must clarify that Lagrangian functions (6) and (27) are not considered equivalent in our derivation. In fact, we only assume that Lagrangian functions (6) and (27) share the same coefficients $\lambda_2$ and $\lambda_4$. Such an assumption has been confirmed by Gorkov's derivation. In his famous article[15], Gorkov has taken into account BCS Hamiltonian including electromagnetic field. Specifically, Gorkov introduced the electromagnetic field $A_\mu$ into BCS Hamiltonian by using "Principle of Local Gauge Invariance"; while, in such a treatment, neither Maxwell term $F_{\mu\nu}F_{\mu\nu}$ nor Chern-Simons term $\varepsilon^{\mu\nu\lambda}A_\mu \partial_\nu A_\lambda$ produce a change in his differential equation for the thermodynamic Green functions. Therefore, the coefficients $\lambda_2$ and $\lambda_4$ will not receive any substantial contributions from Maxwell term or Chern-Simons term. The readers can realize this point by checking the equation (1) in reference 15.

Substituting equation (29) into equation (25) yields:

$$\left( \frac{48 m^{*2} \pi^2 T_c^2}{7\zeta(3) N \varepsilon_F} \right)^{\frac{1}{4-D}} \cdot \rho = \frac{\lambda_0^{\frac{1}{4-D}}}{\sigma_0}. \tag{30}$$

The thin superconducting film can be thought of as a two-dimensional plane, so we can take $D = 2$. Substituting $D = 2$ into equation (30) we finally get the scaling law:

$$T_c = \beta(N) \cdot \rho^{-1}, \tag{31}$$

where $\beta(N) = \ell \cdot N^{\frac{1}{2}}$ and $\ell = \left( \frac{7\zeta(3)\varepsilon_F \lambda_0}{48 \pi^2 m^{*2} \sigma_0^2} \right)^{\frac{1}{2}}$.

Here $\ell$ is a parameter which will be determined by experimental measurements. It must be noticed that $D = 2$ does not contradict the existence of 3-dimensional Chern-Simons term. In fact, here $D = 2$ denotes film material's dimension rather than space's dimension, and it only occurs in the process of renormalization-group analysis. By taking the different values of $D$ we can see how material's dimension affects the critical behavior.

For simplicity, we might as well think of the thin superconducting film as a thin sheet of superconducting material of length and width $l$ and thickness $d \ll l$. As a result, by Zee's method we can get (see page 348 in reference 8):

$$\frac{1}{R_s} = \sigma \cdot d, \tag{32}$$

which by using equation (5) yields:

$$\rho = d \cdot R_s. \tag{33}$$

Substituting equation (33) into equation (31) we get:

$$d \cdot T_c = \beta(N) \cdot R_s^{-1}. \tag{34}$$

The scaling formula (34) is the main result of this paper. It successfully reproduces the scaling law (1). Remarkably, we obtain the theoretical value of critical exponent $B$, which equals 1. We list the experimental value in Table 1.

## Discussion and Conclusion

Although there are a few differences (about 0.05) between theoretical value and experimental mean-value, we stress that the renormalization-group equations (21) and (22) only involve the first-order perturbation. If we consider high order perturbations, the differences may be explained. We hope to explore this point in a more detailed work.

However, we must point out that our formula (34) is somewhat different from the formula (1). This is because the formula (1) indicates that $A$ and $B$ are unknown parameters, but our formula (34) implies that $A$ will depend on the number density of the electrons at the normal state, $N$. This means that we can improve the critical temperature $T_c$ not only by adjusting $d$ and $R_s$, but also by $N$. For example, by our scaling formula (34) the enhancement of $T_c$ will depend on the increase in $N$ and decrease in $d \cdot R_s$. Therefore, we propose to seek some film materials, which are thin enough and meanwhile own high concentration of the electrons and low sheet resistance at the normal state, for manufacturing high-$T_c$ superconducting materials. In addition, we need to clarify that the conductivity $\sigma$ in equation (4) denotes off-diagonal conductivity rather than diagonal conductivity. Thus, our theoretical result (34) can be only regarded as a possible explanation for the experimental result (1), since we do not know whether Ivry, Y. et al have distinguished off-diagonal conductivity and diagonal conductivity in their work[2].

In conclusion, by Wilson's renormalization-group method we have successfully derived the scaling formula (34) near superconducting-to-insulating transition, which has been confirmed by experimental investigations. Our theoretically computed value for critical exponent is in agreement with experiment as well. More importantly, our scaling formula (34) indicates that the superconducting transition temperature $T_c$ will depend mainly on three parameters, which are film thickness ($d$), sheet resistance of the film at the normal state ($R_s$), and number density of the electrons at the normal state ($N$). This finding may provide a possible outlet for facilitating the enhancement of $T_c$.

**Acknowledgments**

This work was supported by the Fundamental Research Funds for the Central Universities (Grant No. SWU1409444) and the National Soft Science Research Plan (Grant No. 2013GXS4D143)

**Additional information**

Competing financial interests: The author declares no competing financial interests.

Table 1: Experimental mean-value[2] and theoretical value of critical exponent

| Critical exponent | Experimental mean-value | Theoretical value |
|---|---|---|
| $B$ | 0.95 | 1 |